\documentclass[reprint,superscriptaddress,amsmath,amssymb,aps,notitlepage,longbibliography,floatfix,nofootinbib,twocolumn,prd]{revtex4-2}

\usepackage{graphicx}
\usepackage{dcolumn}
\usepackage{bm}
\usepackage{natbib}

\usepackage{color}
\usepackage{hyperref}
\hypersetup{colorlinks=true,linkcolor=blue,filecolor=blue,urlcolor=blue,citecolor=blue}
\usepackage{hyperref}

\usepackage{booktabs} 
\usepackage{threeparttable}

\begin{document}%
\begin{sloppypar}

\title{
The Hierarchical Merger Scenario for GW231123
}

\author{Guo-Peng Li}
\affiliation{
Department of Astronomy, School of Physics and Technology, Wuhan University, Wuhan 430072, China}

\author{Xi-Long Fan}
\email[Corresponding author: Xi-Long Fan, ]{xilong.fan@whu.edu.cn}
\affiliation{
Department of Astronomy, School of Physics and Technology, Wuhan University, Wuhan 430072, China}

\date{\today}

\begin{abstract}

GW231123 exhibits exceedingly massive components and high spins, which challenges the formation of first-generation (1G) black holes from stellar collapse and implies that this event might originate from hierarchical mergers. 
Here we show that the \texttt{2G+2G} merger scenario for GW231123 is favored over a \texttt{2G+1G} (or \texttt{3G+2G}) merger, with odds ratios of $>$$\mathcal{O}(10^3)$.
The primary (secondary) black hole is consistent with merging binary black holes with masses of $75^{+7}_{-7}\,M_\odot$ and $66^{+4}_{-10}\,M_\odot$ ($64^{+8}_{-10}\,M_\odot$ and $59^{+7}_{-9}\,M_\odot$; 90\% credible intervals), respectively. 
Our results reveal that the treatment of spin priors from the population level and waveform model choice are decisive in interpreting potential hierarchical gravitational-wave signals.


\end{abstract}

\maketitle

\section{Introduction}

Recently, a remarkable gravitational-wave (GW) event, GW231123, was observed by the Advanced LIGO Hanford and Livingston detectors and reported by the LIGO-Virgo-KAGRA Collaboration~\citep{2025arXiv250708219T}. This event is consistent with the merger of two black holes (BHs) with extremely high masses of $\sim$$137\,M_\odot$ and $\sim$$103\,M_\odot$, which is even rarer than GW190521, with a total mass of $\sim$$140\,M_\odot$~\citep{2020PhRvL.125j1102A}.

Generally, most binary black hole (BBH) mergers can be explained by first-generation (1G) mergers formed from stellar collapse. They primarily arise from isolated binary evolution, as well as dynamical formation in young/open star clusters and globular clusters~\citep{2021hgwa.bookE..16M,2022PhR...955....1M}. Other mergers may originate from second (or higher) generation mergers~\citep{2024PhRvL.133e1401L,2025PhRvL.134a1401A}, referred to as hierarchical (repeated) mergers~\citep{2016ApJ...824L..12O,2017PhRvD..95l4046G,2017ApJ...840L..24F}, with at least one of the BHs being the remnant of a previous merger. These are thought to occur mainly in active galactic nucleus disks and nuclear star clusters~\citep{2021NatAs...5..749G,2023Univ....9..138A}, although they can also occur in globular clusters and young star clusters, where their merger rates strongly depend on the escape speed of the host environment~\citep{2020PhRvD.102l3016A,2019PhRvD.100d3027R,2021MNRAS.505..339M,2021ApJ...915L..35K,2022A&A...666A.194L,2023MNRAS.520.5259A}.

However, the high component masses of GW231123 challenge the formation of 1G BHs through standard stellar collapse due to pair-instability (PI) mass gap ($\sim$$65$–$130\,M_\odot$) predicted by PI supernovae~\citep{2003ApJ...591..288H} and Pulsational PI supernovae~\citep{2007Natur.450..390W,2016A&A...594A..97B}, suggesting that it needs revised models of massive stellar evolution or alternative formation channels~\citep{2025arXiv250819208M}. 
Instead, possible formation scenarios for GW231123 include hierarchical mergers~\citep{2025arXiv250717551L,2025arXiv250813412D}, 
Population III stars~\citep{2025arXiv250801135T}, 
sustained accretion~\citep{2025arXiv250808558B},  
stellar physics (and/or chemically homogeneous evolution)~\citep{2025arXiv250715967S,2025arXiv250815887G,2025arXiv250810088C,2025arXiv250900154P,Baumgarte:2025syh}, 
strong lensing~\citep{2025arXiv250821262S}, 
primordial BHs~\citep{2025arXiv250715701Y,2025arXiv250809965D,2025arXiv250807524N}, 
and cosmic strings~\citep{2025arXiv250720778C}. 
In addition, the two BHs of GW231123 exhibit extremely high spins with dimensionless magnitudes of $\sim$0.9 and $\sim$0.8, which may be consistent with the spins of hierarchical mergers clustering at the characteristic value of $\sim$0.7~\citep{2021NatAs...5..749G}.

In this study, we investigate the properties of GW231123 and its progenitors under the assumption that the event originates from the hierarchical merger scenario. 
We first describe the methodology employed to explore the hierarchical origin of GW231123. Then, we present the results and discussion. Finally, we summarize our conclusions.

\section{Method}\label{sec:method}

We investigate three types of hierarchical BBH mergers for GW231123: 
\texttt{2G+1G}, \texttt{2G+2G}, and \texttt{3G+2G}, corresponding to the three hypotheses 
$\mathcal{H}_{\rm 2G+1G}$, $\mathcal{H}_{\rm 2G+2G}$, and $\mathcal{H}_{\rm 3G+2G}$, respectively.

\subsection{Bayesian Criterion}

To assess which hierarchical merger type is more likely as the formation channel of GW231123, we employ a Bayesian statistical analysis. 
We calculate the \textit{odds ratio} $\mathcal{O}^i_j$ between two hypotheses, $\mathcal{H}_i$ and $\mathcal{H}_j$, defined as
\begin{equation}
    \mathcal{O}^i_j(d) = 
    \frac{p(\mathcal{H}_i|d)}{p(\mathcal{H}_j|d)} =
    \frac{p(d|\mathcal{H}_i)}{p(d|\mathcal{H}_j)}\,
    \frac{p(\mathcal{H}_i)}{p(\mathcal{H}_j)}\,,
    \label{odds ratio}
\end{equation}
where $d$\footnote{\url{https://zenodo.org/records/16004263}} is the GW231123 dataset, 
$\mathcal{B}^i_j = p(d|\mathcal{H}_i)/p(d|\mathcal{H}_j)$ is the \textit{Bayes factor} (i.e., the ratio of evidences), and 
$\mathcal{P}^i_j = p(\mathcal{H}_i)/p(\mathcal{H}_j)$ is the \textit{prior odds}, with $p(\mathcal{H}_i)$ the \textit{prior probability} of $\mathcal{H}_i$.


\subsection{Hypothetical Evidence}

The parameters of interest for this study are the masses, $m_1$ and $m_2$, and spins, $\chi_1$ and $\chi_2$, of the primary and secondary BHs, respectively. 
The evidence $p(d|\mathcal{H}_i)$ is computed by marginalizing over these parameters:
\begin{equation}
\begin{aligned}
    p(d|\mathcal{H}_i) = 
    \int & {\rm d}m_1\, {\rm d}\chi_1\, {\rm d}m_2\, {\rm d}\chi_2\,
    p(d|m_1,\chi_1,m_2,\chi_2) \\
    & \times p(m_1,\chi_1,m_2,\chi_2|\mathcal{H}_i)\,,
    \label{d|Hi;m,chi}
\end{aligned}
\end{equation}
where $p(m_1,\chi_1,m_2,\chi_2|\mathcal{H}_i)$ is the prior conditioned under hypothesis $\mathcal{H}_i$. 
Under different hypotheses, it reads
\begin{equation}
\begin{aligned}
    &p(m_1,\chi_1,m_2,\chi_2|\mathcal{H}_{\rm 2G+1G}) = 
    p(m_1,\chi_1|{\rm 2G})\, p(m_2,\chi_2|{\rm 1G})\,,\\
    &p(m_1,\chi_1,m_2,\chi_2|\mathcal{H}_{\rm 2G+2G}) = 
    p(m_1,\chi_1|{\rm 2G})\, p(m_2,\chi_2|{\rm 2G})\,,\\
    &p(m_1,\chi_1,m_2,\chi_2|\mathcal{H}_{\rm 3G+2G}) = 
    p(m_1,\chi_1|{\rm 3G})\, p(m_2,\chi_2|{\rm 2G})\,,
\end{aligned}
\end{equation}
where $p(m,\chi|N{\rm G})$ represents the mass and spin distribution of the $N$-generation BH population. We note that there is another possibility, such as the case where the primary BH is a 1G BH, while the secondary BH is a 2G BH, but when the mass difference between the two black holes of GW231123 is relatively large, the probability of this situation can be ignored and does not affect our results.

\begin{figure}
\centering
\includegraphics[width=8cm]{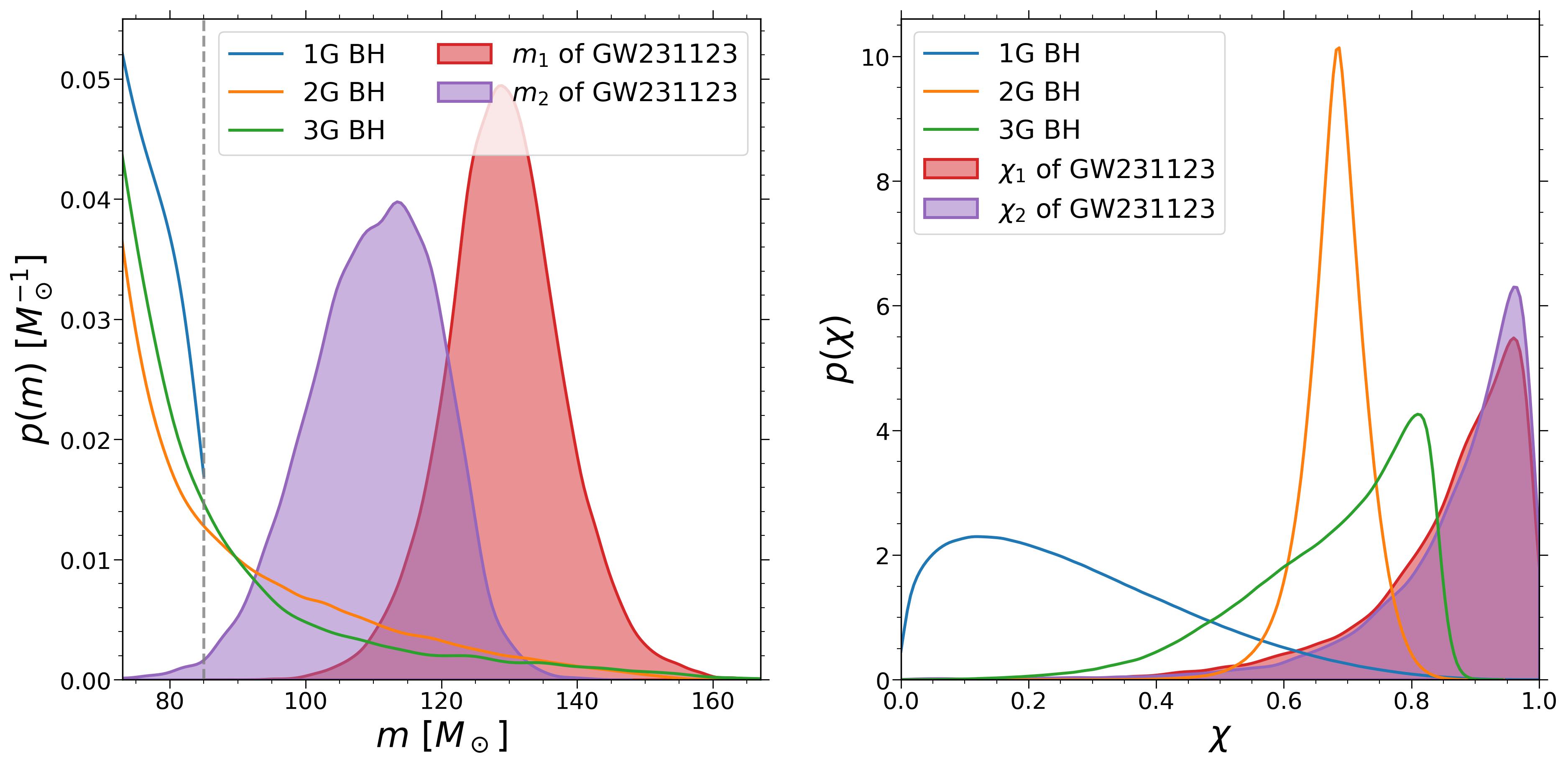}
\caption{
Probability density distributions of masses ($m$, left) and spin magnitudes ($\chi$, right) for 1G (blue), 2G (orange), and 3G (green) BH populations (normalized with $>$$65\,M_\odot$ for visualization). 
Filled red (purple) curves indicate the primary (secondary) BH of GW231123 from the \texttt{NRSur7dq4} waveform model~\citep{2025arXiv250708219T}. 
The 1G BH population follows the GWTC-3 distribution~\citep{2023PhRvX..13a1048A} with a maximum mass of $\sim$$85\,M_\odot$ (vertical gray dashed). 
High-generation BHs use a parametric model of with an escape speed of $\sim$$100\,{\rm km\,s^{-1}}$~\citep{2023PhRvD.107f3007L,2025ApJ...981..177L} and isotropic spin tilts.
}
\label{fig1} 
\end{figure}

Figure~\ref{fig1} shows the distribution characteristics of 1G, 2G, and 3G BHs in terms of mass and spin, compared with GW231123 for visualization. 
Within the mass range of GW231123, the 1G BH mass distribution declines rapidly and truncates at a maximum mass of $\sim$$85\,M_\odot$, indicating that the fraction of 1G BHs decreases quickly with increasing mass and the probability at high masses (e.g., within the GW231123 range) is extremely low. In contrast, the mass distribution peaks of 2G and 3G BHs are located closer to the high-mass region, with maximum masses of $\sim$$158\,M_\odot$ and $\sim$$197\,M_\odot$, respectively. The primary and secondary masses, $m_1$ and $m_2$, of GW231123 mainly lie in the range $\sim$$100$–$140\,M_\odot$, overlapping with the mass distributions of 2G and 3G BHs. The match is higher for 2G BHs, although 3G BHs extend to larger masses.
Regarding spin, the 1G BH spin distribution peaks at low values ($\chi\sim0$–$0.2$) and decreases rapidly, indicating a tendency toward low-spin states. The spin distributions of 2G and 3G BHs peak at higher spins ($\chi\sim0.7$–$0.9$), with the 3G distribution being relatively flat. The observed spins, $\chi_1$ and $\chi_2$, of GW231123 are mainly in the high-spin range ($\chi\sim0.7$–$1$), showing a good agreement with the 3G BH spin distribution, whereas 1G BHs have a poor match due to their low-spin tendency.
We note that in computing the evidence to infer the progenitor properties of GW231123, we do not directly use the 2G and 3G BH population distributions (see Figure~\ref{fig1}). Instead, they are synthesized individually using the parametric population model~\citep{2023PhRvD.107f3007L,2025ApJ...981..177L}, starting from the 1G BH population~\citep{2023PhRvX..13a1048A}. During this process, the escape speed of the host environment is used to constrain the corresponding parameter space, implying that the progenitor merger should have as equal-mass binaries and small spins as possible; otherwise, the remnant BH may be ejected due to a kick velocity exceeding the escape speed (see Appendix~\ref{appendix1}).

\subsection{Prior Probability}

We calculate the prior probability $p(\mathcal{H}_i)$ by setting it proportional to the retention probability, $f_{\rm ret}$. This is motivated by the fact that a primary condition for hierarchical mergers is that the remnant BHs can be retained by their host~\citep{2022A&A...666A.194L,2025ApJ...984...63L}; thus, the retention probability of progenitor mergers largely reflects the likelihood of future mergers. 
In particular, a \texttt{2G+1G} merger requires a 2G BH formed from a \texttt{1G+1G} merger;
similarly, both progenitors of a \texttt{2G+2G} merger originate from independent \texttt{1G+1G} mergers;
for a \texttt{3G+2G} merger, the 2G BH arises from a \texttt{1G+1G} merger, and the 3G BH arises from a \texttt{2G+1G} merger, in which the 2G BH itself comes from another \texttt{1G+1G} merger. Hence, the priors are
\begin{equation}
\begin{aligned}
    p(\mathcal{H}_{\rm 2G+1G}) &\propto f^{\rm 1G+1G}_{\rm ret}\,,\\
    p(\mathcal{H}_{\rm 2G+2G}) &\propto f^{\rm 1G+1G}_{\rm ret} \times f^{\rm 1G+1G}_{\rm ret}\,,\\
    p(\mathcal{H}_{\rm 3G+2G}) &\propto
    f^{\rm 2G+1G}_{\rm ret} \times f^{\rm 1G+1G}_{\rm ret} \times f^{\rm 1G+1G}_{\rm ret}\,,
    \end{aligned}
\end{equation}
where $f^{\rm 1G+1G}_{\rm ret}$ ($f^{\rm 2G+1G}_{\rm ret}$) is the retention probability of \texttt{1G+1G} (\texttt{2G+1G}) mergers. We note that we neglect the delay times between subsequent mergers~\citep{2012ApJ...759...52D,2017ApJ...835..165B,2020MNRAS.497.1043D,2021Symm...13.1678M}, which may also affect hierarchical merger rates. This assumption is justified because the effect of delay times is negligible in environments with relatively low escape speeds, such as globular and nuclear star clusters~\citep{2023PhRvD.107f3007L}.

\begin{figure}
\centering
\includegraphics[width=8cm]{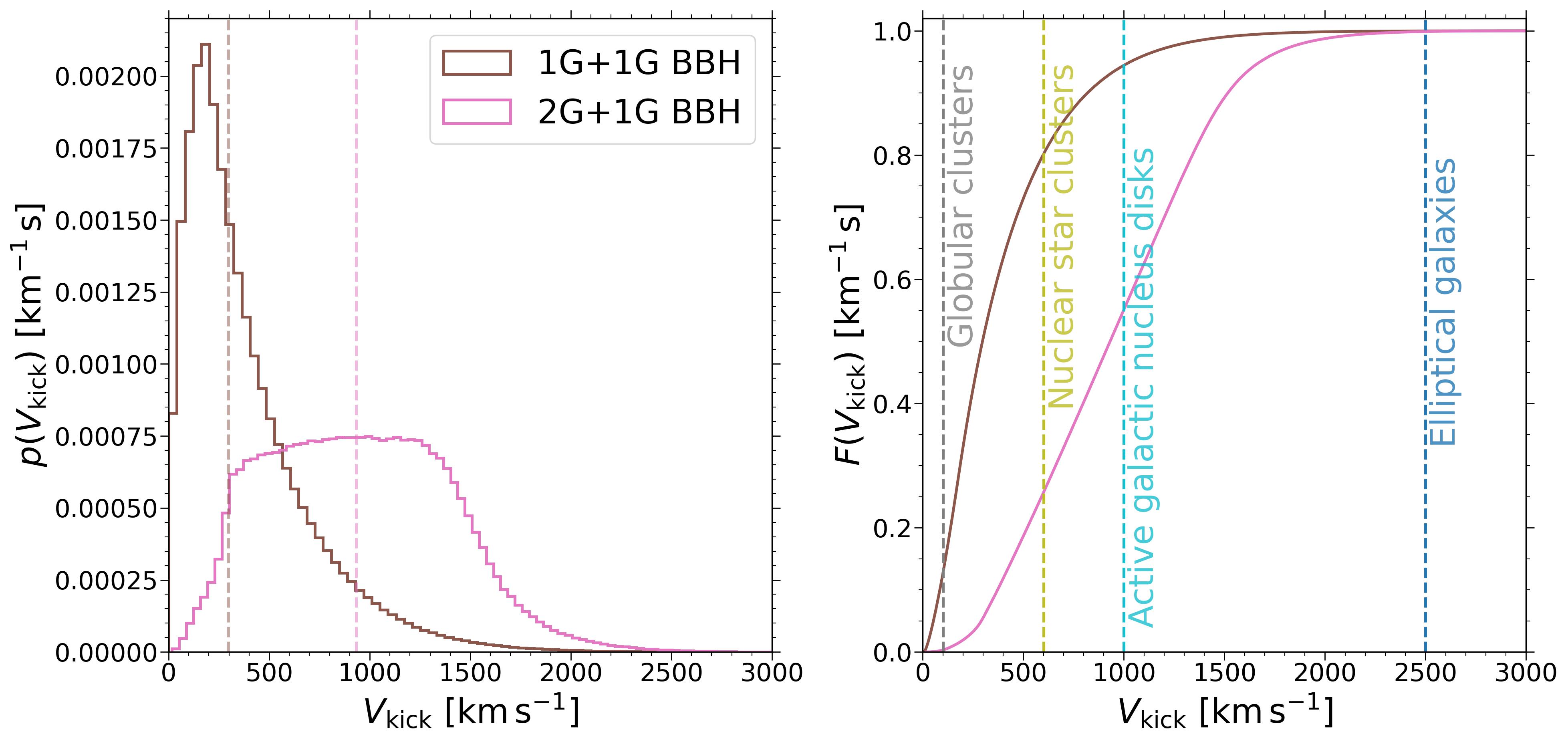}
\caption{
Probability density distributions (left) of kick velocities ($V_{\rm kick}$) for \texttt{1G+1G} (brown) and \texttt{2G+1G} (pink) BBH mergers, along with their cumulative distributions (right). Vertical dashed lines indicate the 50th percentiles of $V_{\rm kick}$ (left) and the upper bounds of host environment escape speeds (right) for comparison.
}
\label{fig2} 
\end{figure}

Figure~\ref{fig2} shows the probability density and cumulative (retention probability) distributions of kick velocities ($V_{\rm kick}$) for \texttt{1G+1G} and \texttt{2G+1G} mergers, spanning a wide range of $\sim$$0$–$4000\,{\rm km\,s^{-1}}$. 
The probability distribution of \texttt{1G+1G} kick velocities peaks at relatively low values ($\sim$$0$–$500\,{\rm km\,s^{-1}}$) and declines rapidly with increasing $V_{\rm kick}$. In contrast, \texttt{2G+1G} mergers peak at higher $V_{\rm kick}$ ($\sim$$500$–$1500\,{\rm km\,s^{-1}}$). The median kick velocities are $297_{-244}^{+734}\,{\rm km\,s^{-1}}$ and $931_{-640}^{+745}\,{\rm km\,s^{-1}}$ (90\% credible intervals) for \texttt{1G+1G} and \texttt{2G+1G} mergers, respectively; the higher values for \texttt{2G+1G} arise from their larger spins and lower mass ratios. 
The cumulative distributions of $V_{\rm kick}$ rise more steeply at low velocities for \texttt{1G+1G} mergers, whereas those of \texttt{2G+1G} extend to higher velocities. For comparison, the right panel shows dashed lines marking the upper bounds of typical escape speed ranges for different environments from $\mathcal{O}(10)\,{\rm km\,s^{-1}}$ for globular clusters~\citep{2016ApJ...831..187A} to $\mathcal{O}(100)\,{\rm km\,s^{-1}}$ for nuclear star clusters~\citep{2016ApJ...831..187A} and up to $\mathcal{O}(1000)\,{\rm km\,s^{-1}}$ in active galactic nucleus disks. Accordingly, the retention probability of \texttt{1G+1G} mergers is $\sim$0.13 for an escape speed of $100\,{\rm km\,s^{-1}}$, while that of \texttt{2G+1G} mergers is only $\sim$$1.55\times10^{-5}$. This indicates that \texttt{1G+1G} mergers are more compatible with low escape speed environments such as globular clusters, whereas \texttt{2G+1G} mergers preferentially match higher escape speed environments, including nuclear star clusters and active galactic nucleus disks, suggesting that hierarchical mergers contribute to producing BBHs across a broader range of cosmic settings.

\section{Results and Discussion}\label{sec:RD}

We first compute the prior odds, Bayes factors, and odds ratios of GW231123 among \texttt{2G+1G}, \texttt{2G+2G}, and \texttt{3G+2G} mergers, and then analyze the progenitor properties under the most likely hierarchical merger scenario.

\subsection{Is GW231123 a \texttt{2G+2G} merger?}

\begin{table*}
\centering
\caption{Model comparison results.}
\label{tab}
\begin{tabular}{ccccccccc}
\toprule
$V_{\rm esc}\,[{\rm km\,s^{-1}}]$ & $M_{\rm max}\,[M_\odot]$ & waveform & $\mathcal{P}^{\rm 2G+2G}_{\rm 2G+1G}$ & $\mathcal{P}^{\rm 2G+2G}_{\rm 3G+2G}$ & ${\rm log}\mathcal{B}^{\rm 2G+2G}_{\rm 2G+1G}$ & ${\rm log}\mathcal{B}^{\rm 2G+2G}_{\rm 3G+2G}$ & $\mathcal{O}^{\rm 2G+2G}_{\rm 2G+1G}$ & $\mathcal{O}^{\rm 2G+2G}_{\rm 3G+2G}$ \\
\midrule
100 & 85 & \texttt{NRSur7dq4} & 0.13 & $6.46\times10^4$ & 11.53 & 8.38 & $1.28\times10^4$ & $2.81\times10^8$ \\
300 & 85 & \texttt{NRSur7dq4} & 0.56 & $2.48\times10^4$ & 7.61 & 1.66 & 1131 & $1.31\times10^5$ \\
100 & 65 & \texttt{NRSur7dq4} & 0.20 & $1.76\times10^5$ & 13.46 & 9.04 & $1.37\times10^5$ & $1.49\times10^9$ \\
100 & 85 & \texttt{IMRPhenomXO4a} & 0.13 & $6.46\times10^4$ & $-1.48$ & / & $0.03$ & / \\
\bottomrule
\end{tabular}
\begin{tablenotes} 
\item 
{\bf Note.}
Column (1): escape speed of the host environment.  
Column (2): maximum mass of 1G BHs.  
Column (3): waveform model used to derive the posterior distribution of GW231123.  
Columns (4-9): results of the prior odds, Bayes factors, and odds ratios among \texttt{2G+1G}, \texttt{2G+2G}, and \texttt{3G+2G} mergers. 
``/'' indicates that the relevant parameter space could not be explored due to computational cost.
\end{tablenotes}
\end{table*}

Table~\ref{tab} presents model comparison results among the three hypotheses $\mathcal{H}_{\rm 2G+1G}$, $\mathcal{H}_{\rm 2G+2G}$, and $\mathcal{H}_{\rm 3G+2G}$ for GW231123, under different assumptions for the host environment escape speed, maximum mass of 1G BHs, and waveform model. 
Using the \texttt{NRSur7dq4} waveform model~\citep{2019PhRvR...1c3015V}, for $V_{\rm esc}=100\,{\rm km\,s^{-1}}$ and $M_{\rm max}=85\,M_\odot$, the prior odds are $\mathcal{P}^{\rm 2G+2G}_{\rm 2G+1G}=0.13$ and $\mathcal{P}^{\rm 2G+2G}_{\rm 3G+2G}=6.46 \times 10^4$, indicating a much higher prior likelihood for \texttt{2G+1G} mergers relative to \texttt{3G+2G} when compared to \texttt{2G+2G}. The Bayes factors are ${\rm log}\mathcal{B}^{\rm 2G+2G}_{\rm 2G+1G}=11.53$ and ${\rm log}\mathcal{B}^{\rm 2G+2G}_{\rm 3G+2G}=8.38$, showing strong evidence favoring \texttt{2G+2G} over both \texttt{2G+1G} and \texttt{3G+2G}, with a stronger preference over \texttt{2G+1G}. The corresponding odds ratios are $\mathcal{O}^{\rm 2G+2G}_{\rm 2G+1G}=1.28 \times 10^4$ and $\mathcal{O}^{\rm 2G+2G}_{\rm 3G+2G}=2.81 \times 10^8$, demonstrating a vast difference in likelihood between these merger scenarios. This indicates that the most probable hierarchical merger scenario for GW231123 is \texttt{2G+2G}.
For a higher escape speed, $V_{\rm esc}=300\,{\rm km\,s^{-1}}$ (still \texttt{NRSur7dq4} and $M_{\rm max}=85\,M_\odot$), the prior odds change to $\mathcal{P}^{\rm 2G+2G}_{\rm 2G+1G}=0.56$ and $\mathcal{P}^{\rm 2G+2G}_{\rm 3G+2G}=2.48 \times 10^4$, indicating that higher escape speeds lower the threshold for remnant BHs to be retained, making higher-generation mergers more likely. In this case, ${\rm log}\mathcal{B}^{\rm 2G+2G}_{\rm 2G+1G}=7.61$, $\mathcal{O}^{\rm 2G+2G}_{\rm 2G+1G}=1131$, ${\rm log}\mathcal{B}^{\rm 2G+2G}_{\rm 3G+2G}=1.66$, and $\mathcal{O}^{\rm 2G+2G}_{\rm 3G+2G}=1.31 \times 10^5$, showing that the relative preference for \texttt{2G+2G} decreases with increasing escape speed, although it remains the favored scenario over \texttt{2G+1G} and \texttt{3G+2G}.
To ensure that the results are robust against the location of the theorized PI mass gap, we repeated the analysis with $M_{\rm max}=65\,M_\odot$. We find that a lower maximum mass for 1G BHs significantly increases both the odds and prior likelihood for \texttt{2G+2G} relative to \texttt{2G+1G} and \texttt{3G+2G}, and strengthens the evidence supporting \texttt{2G+2G} as the preferred merger scenario.

Our analysis above uses the posterior distribution of GW231123 obtained with the \texttt{NRSur7dq4} waveform model~\citep{2019PhRvR...1c3015V}. However, the mass distribution from the \texttt{IMRPhenomXO4a} waveform~\citep{2024PhRvD.109f3012T} differs significantly: the primary mass is larger ($\sim$$143\,M_\odot$ vs. $\sim$$129\,M_\odot$) and the secondary mass is smaller ($\sim$$55\,M_\odot$ vs. $\sim$$111\,M_\odot$)~\citep{2025arXiv250708219T}. 
Under this alternative waveform, we compute ${\rm log}\mathcal{B}^{\rm 2G+2G}_{\rm 2G+1G}=-1.48$ and $\mathcal{O}^{\rm 2G+2G}_{\rm 2G+1G}=0.03$. This indicates that the probability of the \texttt{2G+2G} hypothesis is only $\sim$3\% of the \texttt{2G+1G} hypothesis, providing evidence against \texttt{2G+2G} relative to \texttt{2G+1G}. 
The large shift in relative probability highlights the critical sensitivity of BBH merger inferences to the choice of waveform model. The stark difference in mass distributions between \texttt{NRSur7dq4} and \texttt{IMRPhenomXO4a} (e.g., $m_2$ decreasing from $\sim$$111\,M_\odot$ to $\sim$$55\,M_\odot$) directly affects the compatibility of GW231123 with hierarchical merger scenarios, emphasizing that waveform selection is not a trivial technical detail but a key factor shaping conclusions about BH origin channels, particularly for unusually massive events like GW231123.
Furthermore, analysis of the \texttt{3G+2G} hypothesis is extremely challenging due to the difficulty in exploring the relevant progenitor parameter space, illustrating that higher-generation mergers (and the choice of waveform model) can strongly influence the feasibility and outcomes of hierarchical merger model comparisons.

\subsection{Progenitors of GW231123}

\begin{figure}
\centering
\includegraphics[width=8cm]{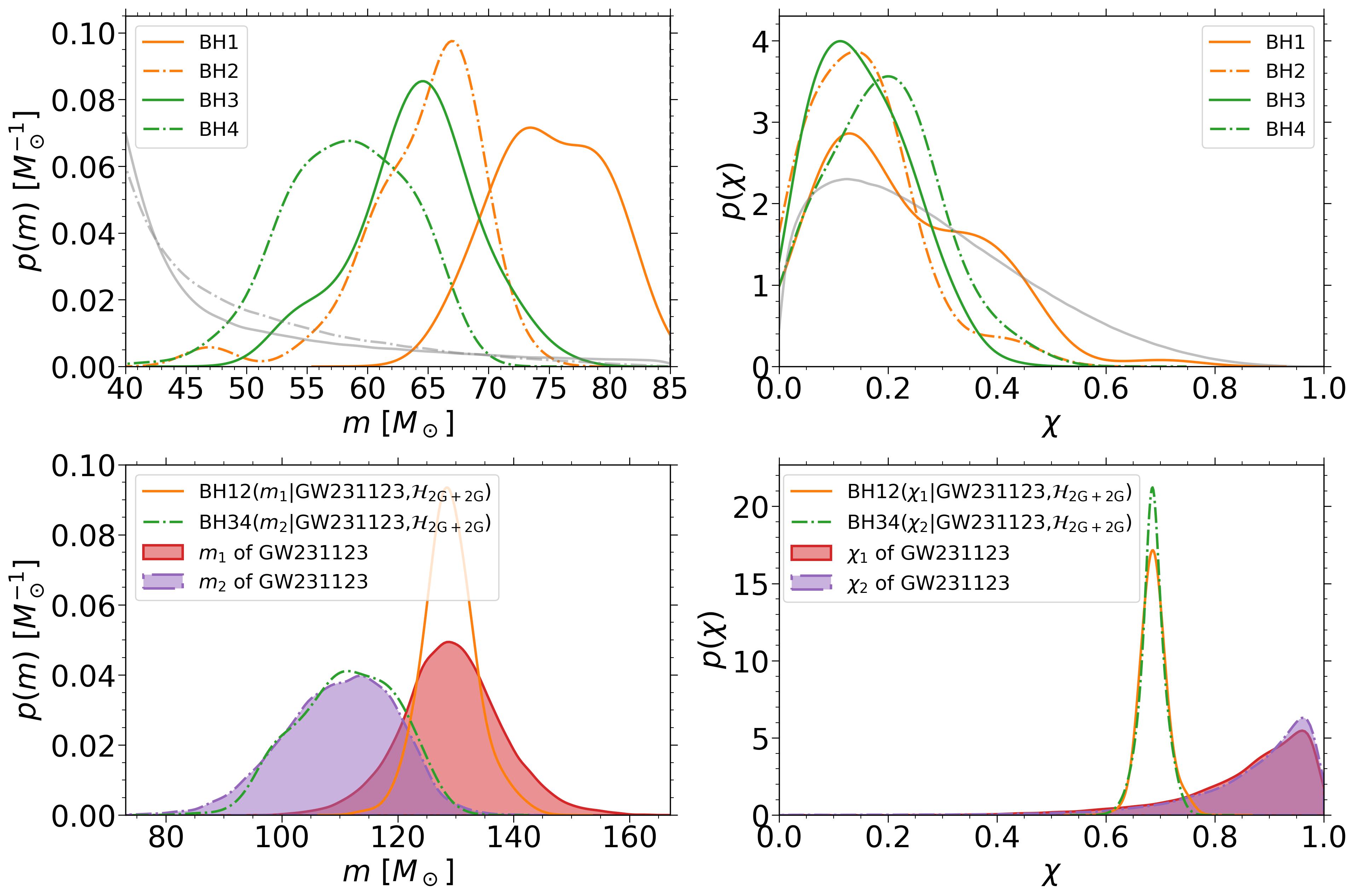}
\caption{
Posterior probability density distributions of masses and spins for progenitor BHs of GW231123, compared with the primary and secondary BHs. 
\textit{Top panels:} posterior distributions of masses ($m$, left) and spin magnitudes ($\chi$, right) of the four 1G BHs that merged hierarchically to form GW231123. The orange (BH1 and BH2) and green (BH3 and BH4) curves correspond to the progenitor mergers forming the primary and secondary BHs of GW231123, with solid lines representing the primary BHs (BH1 and BH3) and dot-dashed lines representing the secondary BHs (BH2 and BH4), of progenitor mergers. The gray solid (dashed) curve shows the prior distribution of the primary (secondary) BHs of \texttt{1G+1G} mergers (normalized with $m>35\,M_\odot$ for visualization). The spin priors for primary and secondary BHs are identical. 
\textit{Bottom panels:} posterior distributions of masses (left) and spin magnitudes (right) of the two 2G BHs resulting from the mergers of the four 1G BHs. The orange solid (green dot-dashed) curve represents the primary (secondary) BH of GW231123, while the filled red (purple) curve shows the primary (secondary) BH from the \texttt{NRSur7dq4} waveform model~\citep{2025arXiv250708219T} for comparison.
}
\label{fig3} 
\end{figure}

Figure~\ref{fig3} illustrates the posterior probability distributions of masses and spins for the progenitor BHs of GW231123, assuming the event is most likely a \texttt{2G+2G} merger based on the previous analysis. 
The progenitor BHs, labeled BH1, BH2, BH3, and BH4, merge hierarchically such that BH1 and BH2 form the primary BH (BH12) of GW231123, while BH3 and BH4 form the secondary BH (BH34). In the first progenitor merger (BH1 + BH2) forming the primary BH, both BHs have relatively high masses of $75^{+7}_{-7}\,M_\odot$ and $66^{+4}_{-10}\,M_\odot$, producing a remnant BH with a final mass of $129^{+8}_{-7}\,M_\odot$, which is more tightly constrained than the primary BH mass of $129^{+15}_{-14}\,M_\odot$ from the \texttt{NRSur7dq4} waveform model. In the second progenitor merger (BH3 + BH4) forming the secondary BH, the BHs have relatively lower masses of $64^{+8}_{-10}\,M_\odot$ and $59^{+7}_{-9}\,M_\odot$, resulting in a remnant BH with a final mass of $112^{+13}_{-14}\,M_\odot$, compared to the secondary BH mass of $111^{+14}_{-17}\,M_\odot$ from \texttt{NRSur7dq4}. These results suggest that these hierarchical mergers are plausible progenitors and provide insights into the formation and properties of the GW231123 event.

For the four progenitor BHs of GW231123, the spin magnitude distributions all peak at relatively low values ($\sim$0.1--0.3), with medians and 90\% credible intervals of $0.19_{-0.15}^{+0.28}$, $0.14_{-0.12}^{+0.22}$, $0.14_{-0.11}^{+0.17}$, and $0.19_{-0.16}^{+0.18}$, respectively. This low-spin result arises because larger spins lead to higher kick velocities, which can eject the BHs from their host environment, preventing them from participating in subsequent mergers. The two progenitor mergers of GW231123 both show post-merger spin distributions sharply peaked near $\chi \sim 0.7$, with $0.69_{-0.05}^{+0.04}$ and $0.69_{-0.04}^{+0.04}$, reflecting the effect of the spin prior that clusters around 0.7~\citep{2021NatAs...5..749G}. In contrast, the spins of GW231123 from the \texttt{NRSur7dq4} waveform are clustered at higher values ($\chi\sim0.7$--1) under the assumption of a uniform prior in $[0,1]$ during parameter estimation. This highlights the importance of reconsidering the spin prior distribution for high-generation BHs, which tend to cluster around $\sim$0.7, rather than assuming a uniform distribution, when estimating parameters for potential hierarchical merger GW events.

\section{Conclusions}\label{sec:conclusions}

We investigate the origin of the exceptionally massive BBH merger signal GW231123~\citep{2025arXiv250708219T}, assuming this event is a hierarchical merger, as its properties challenge the formation of BHs from standard stellar collapse due to pair instability mechanisms. We consider three types of hierarchical mergers for GW231123: \texttt{2G+1G}, \texttt{2G+2G}, and \texttt{3G+2G}. We further examine the progenitor properties of GW231123 under the most likely hierarchical scenario.
Our findings are summarized as follows: 
(1) GW231123 is most likely a \texttt{2G+2G} merger. In particular,  by taking into account the masses and spin magnitudes of the source with the \texttt{NRSur7dq4} waveform model and setting $V_{\rm esc}=100\,{\rm km\,s^{-1}}$ and $M_{\rm max}=85\,M_\odot$, we find that the \texttt{2G+2G} scenario is strongly favored over \texttt{2G+1G} (\texttt{3G+2G}) mergers, with log Bayes factors of 11.53 (8.35), corresponding to odds ratios of $\sim$$1.28\times10^4~(2.81\times10^8):1$, assuming astrophysical prior odds of 0.13 ($6.46\times10^4$).
(2) The primary (secondary) BH of GW231123 likely originates from a progenitor merger with component masses of $75^{+7}_{-7}\,M_\odot$ and $66^{+4}_{-10}\,M_\odot$ ($64^{+8}_{-10}\,M_\odot$ and $59^{+7}_{-9}\,M_\odot$), resulting in remnant masses of $129^{+8}_{-7}\,M_\odot$ and $112^{+13}_{-14}\,M_\odot$, which are consistent with the masses of $129^{+15}_{-14}\,M_\odot$ and $111^{+14}_{-17}\,M_\odot$ of GW231123, respectively, from the \texttt{NRSur7dq4} waveform.
(3) The four progenitor BHs have relatively low spins ($\sim$0–0.4), producing remnant BHs with spins sharply peaked near $\sim$0.7, dominated by the prior reflecting the clustering of hierarchical merger spins at the characteristic value of $\sim$0.7. This indicates that spin priors should be reconsidered, rather than assuming a uniform distribution, when estimating parameters of potential hierarchical merger GW signals.
(4) Changing the waveform from \texttt{NRSur7dq4} to \texttt{IMRPhenomXO4a}, which significantly alters the mass distributions, drastically reduces the \texttt{2G+2G} hypothesis probability to only $\sim$3\% of the \texttt{2G+1G} hypothesis. This underscores that waveform choice is a critical, non-trivial factor in shaping conclusions about BH merger origin channels for such massive events.

\section{Acknowledgments}
This work is supported by National Key R$\&$D Program of China (2020YFC2201400). 
This research has made use of data or software obtained from the Gravitational Wave Open Science Center (\url{https://gwosc.org}), a service of the LIGO Scientific Collaboration, the Virgo Collaboration, and KAGRA. This material is based upon work supported by NSF's LIGO Laboratory which is a major facility fully funded by the National Science Foundation, as well as the Science and Technology Facilities Council (STFC) of the United Kingdom, the Max-Planck-Society (MPS), and the State of Niedersachsen/Germany for support of the construction of Advanced LIGO and construction and operation of the GEO600 detector. Additional support for Advanced LIGO was provided by the Australian Research Council. Virgo is funded, through the European Gravitational Observatory (EGO), by the French Centre National de Recherche Scientifique (CNRS), the Italian Istituto Nazionale di Fisica Nucleare (INFN) and the Dutch Nikhef, with contributions by institutions from Belgium, Germany, Greece, Hungary, Ireland, Japan, Monaco, Poland, Portugal, Spain. KAGRA is supported by Ministry of Education, Culture, Sports, Science and Technology (MEXT), Japan Society for the Promotion of Science (JSPS) in Japan; National Research Foundation (NRF) and Ministry of Science and ICT (MSIT) in Korea; Academia Sinica (AS) and National Science and Technology Council (NSTC) in Taiwan.
This analysis was made possible following software packages:
NumPy~\citep{harris2020array}, 
SciPy~\citep{2020SciPy-NMeth}, 
Matplotlib~\citep{2007CSE.....9...90H}, 
emcee~\citep{2013PASP..125..306F}, 
IPython~\citep{2007CSE.....9c..21P},
corner~\citep{2016JOSS....1...24F},
seaborn~\citep{Waskom2021},
and Astropy~\citep{2022ApJ...935..167A}.

\appendix

\section{Priors Conditioned}\label{appendix1}

Here, we introduce the priors conditioned on higher-generation BHs, $p(m,\chi|{\rm 2G})$ and $p(m,\chi|{\rm 3G})$.  
A 2G BH is the remnant of a merger between two 1G BHs, so its prior is given by
\begin{equation}
\begin{aligned}
    p&(m,\chi|{\rm 2G})\\
    &= \int {\rm d}{m^{\rm 1G}_{1}}{\rm d}{\chi^{\rm 1G}_{1}}{\rm d}{m^{\rm 1G}_{2}}{\rm d}{\chi^{\rm 1G}_{2}}\, 
    p(m,\chi|M^{\rm 2G}_{\rm f},\chi^{\rm 2G}_{\rm f})\\
    &~~~~~~\times
    p(M^{\rm 2G}_{\rm f},\chi^{\rm 2G}_{\rm f}|m^{\rm 1G}_{1},\chi^{\rm 1G}_{1},m^{\rm 1G}_{2},\chi^{\rm 1G}_{2})\,\\
    &~~~~~~\times
    \mathcal{H}\big[V_{\rm kick}(m^{\rm 1G}_{1},\chi^{\rm 1G}_{1},m^{\rm 1G}_{2},\chi^{\rm 1G}_{2})-V_{\rm esc}\big]\\
    &~~~~~~\times
    p(m^{\rm 1G}_{1},\chi^{\rm 1G}_{1},m^{\rm 1G}_{2},\chi^{\rm 1G}_{2}|\mathcal{H}_{\rm 1G+1G})\,,
\end{aligned}
\end{equation}
where $\mathcal{H}_{\rm 1G+1G}$ denotes the 1G BBH mergers following the distribution from GWTC-3~\citep{2023PhRvX..13a1048A}; 
$\mathcal{H}$ is the Heaviside step function;  
$M^{\rm 2G}_{\rm f}$ and $\chi^{\rm 2G}_{\rm f}$ are the mass and spin of the 2G BH remnant formed from merging two 1G BHs with masses $m^{\rm 1G}_{1,2}$ and spins $\chi^{\rm 1G}_{1,2}$, respectively;  
and $V_{\rm kick}$ is the kick velocity of the remnant BH, which depends only on the intrinsic parameters of the progenitors.
A critical condition for hierarchical mergers is the retention of remnant BHs in their host environments. This requires that the kick velocities imparted during the previous merger must be smaller than the escape speed of the host environment. Here, the escape speed $V_{\rm esc}$ is defined as the minimum velocity required for an object to escape the gravitational potential of its host and varies significantly across different astrophysical environments. Accordingly, we constrain the parameter space of the progenitors of GW231123 by requiring $V_{\rm kick} < V_{\rm esc}$.

A 3G BH is the remnant of sequence mergers involving three 1G BHs. Its prior conditioned can be expressed as
\begin{equation}
\begin{aligned}
    p&(m,\chi|{\rm 3G})\\
    &= \int {\rm d}m^{\rm 1G}_{11}\,{\rm d}\chi^{\rm 1G}_{11}\,{\rm d}m^{\rm 1G}_{12}\,{\rm d}\chi^{\rm 1G}_{12}\,{\rm d}m^{\rm 1G}_{2}\,{\rm d}\chi^{\rm 1G}_{2}\,\\
    &~~~~~~\times
    p(m,\chi|M^{\rm 3G}_{\rm f},\chi^{\rm 3G}_{\rm f})\\
    &~~~~~~\times
    p(M^{\rm 3G}_{\rm f},\chi^{\rm 3G}_{\rm f}|m^{\rm 2G}_{1},\chi^{\rm 2G}_{1},m^{\rm 1G}_{2},\chi^{\rm 1G}_{2})\,\\
    &~~~~~~\times
    \mathcal{H}\big[V_{\rm kick}(m^{\rm 2G}_{1},\chi^{\rm 2G}_{1},m^{\rm 1G}_{2},\chi^{\rm 1G}_{2})-V_{\rm esc}\big]\\
    &~~~~~~\times
    p(m^{\rm 2G}_{1},\chi^{\rm 2G}_{1}|M^{\rm 2G}_{\rm f},\chi^{\rm 2G}_{\rm f})\,\\
    &~~~~~~\times
    p(M^{\rm 2G}_{\rm f},\chi^{\rm 2G}_{\rm f}|m^{\rm 1G}_{11},\chi^{\rm 1G}_{11},m^{\rm 1G}_{12},\chi^{\rm 1G}_{12})\,\\
    &~~~~~~\times
    \mathcal{H}\big[V_{\rm kick}(m^{\rm 1G}_{11},\chi^{\rm 1G}_{11},m^{\rm 1G}_{12},\chi^{\rm 1G}_{12})-V_{\rm esc}\big]\\
    &~~~~~~\times
    p(m^{\rm 1G}_{11},\chi^{\rm 1G}_{11},m^{\rm 1G}_{12},\chi^{\rm 1G}_{12}|\mathcal{H}_{\rm 1G+1G})\,
    p(m^{\rm 1G}_{2},\chi^{\rm 1G}_{2}|{\rm 1G})\,.
\end{aligned}
\end{equation}
Here, the first merger of a 1G BBH with masses $m^{\rm 1G}_{11,12}$ and spins $\chi^{\rm 1G}_{11,12}$ produces a 2G remnant with mass $M^{\rm 2G}_{\rm f}$ and spin $\chi^{\rm 2G}_{\rm f}$, i.e., $m^{\rm 2G}_{1}$ and $\chi^{\rm 2G}_{1}$.  
Then, this 2G remnant merges with a second 1G BH of mass $m^{\rm 1G}_{2}$ and spin $\chi^{\rm 1G}_2$ to form the 3G BH with final mass $M^{\rm 3G}_{\rm f}$ and spin $\chi^{\rm 3G}_{\rm f}$, i.e., $m$ and $\chi$.  
The Heaviside functions enforce that each remnant is retained in its host environment, i.e., $V_{\rm kick} < V_{\rm esc}$.

\newcommand{\jcap}{J. Cosmol. Astropart. Phys.}
\newcommand{\physrep}{Phys. Rep.}
\newcommand{\mnras}{Mon. Not. R. Astron. Soc.}
\newcommand{\araa}{Annu. Rev. Astron. Astrophys.}
\newcommand{\aap}{Astron. Astrophys}
\newcommand{\aj}{Astron. J.}
\newcommand{\plb}{Phys. Lett. B}
\newcommand{\apjs}{Astrophys. J. Suppl.}
\newcommand{\app} {Astropart. Phys.}
\newcommand{\apjl}{Astrophys. J. Lett}
\newcommand{\pasp}{Publ. Astron. Soc. Pac.}


%

\end{sloppypar}\end{document}